\documentclass[runningheads]{llncs}
\usepackage{amsmath,amssymb,enumerate,amsfonts}
\usepackage{float}
\usepackage{xcolor}

\usepackage{hyperref}

\usepackage{mathtools}
\usepackage{wrapfig}
\usepackage[binary-units=true]{siunitx}
\usepackage{listings}
\usepackage[labelsep=period,labelfont={bf}]{caption}
\usepackage{subcaption}

\usepackage{lipsum}
\usepackage{xpatch}

\usepackage{tikz} 
  \usetikzlibrary{shapes,arrows,positioning,calc,fit,arrows.meta} 
\usepackage{tkz-graph}

\newcommand{\dcocpp}{{\ttfamily dco\kern-.08em{\raisebox{-.1ex}{/}\kern-.15em {c\kern-.03em{\raisebox{-.18ex}{+\kern-.028em{+}}}}}}}

\newcommand{\cpp}{{\ttfamily
{C\kern-.03em{\raisebox{-.18ex}{+\kern-.028em{+}}}}}}

\newcommand{\dcocppeigen}{{\ttfamily
dco\kern-.08em{\raisebox{-.1ex}{/}\kern-.15em
{c\kern-.03em{\raisebox{-.18ex}{+\kern-.028em{+}}}}}\kern-.08em{\raisebox{-.1ex}{/}}eigen}}

\definecolor{mybg}{rgb}{0.99,0.99,0.99}
\definecolor{mygreen}{rgb}{0,0.7,0}
\definecolor{mygray}{rgb}{0.5,0.5,0.5}
\definecolor{mymauve}{rgb}{0.8,0.2,0.5}

\title{Eigen-AD: Algorithmic Differentiation of the Eigen Library}
\author{Patrick Peltzer \and Johannes Lotz \and Uwe Naumann}
\authorrunning{P. Peltzer et al.}
\institute{Informatik 12: Software and Tools for Computational Engineering, RWTH Aachen University, 52056 Aachen, Germany; Email: \email{info@stce.rwth-aachen.de}}

\makeatletter
\newinsert\copyrightnote@ins
\count\copyrightnote@ins=1000
\dimen\copyrightnote@ins=8in
\newcommand*\copyrightnote@hook
  {%
    \unvbox\copyrightnote@ins
    \global\let\@makecol\copyrightnote@makecol
  }
\let\copyrightnote@AtBeginDocument\AtBeginDocument
\AtBeginDocument{\let\copyrightnote@AtBeginDocument\@firstofone}
\newcommand*\copyrightnote@firstuse
  {%
    \gdef\copyrightnote@firstuse
      {\global\skip\copyrightnote@ins=\bigskipamount\gdef\copyrightnote@firstuse{}}%
    \global\let\copyrightnote@makecol\@makecol
    \xpatchcmd\@makecol{\unvbox\footins}{\unvbox\footins\copyrightnote@hook}
      {}{\GenericError{}{patching @makecol failed}{}{}}
    \copyrightnote@AtBeginDocument
      {%
        \ifvoid\footins
          \insert\footins{}%
        \else
          \global\skip\copyrightnote@ins=\bigskipamount
        \fi
      }%
  }
\newcommand\copyrightnote[1]
  {%
    \copyrightnote@firstuse
    \copyrightnote@AtBeginDocument
      {%
        \insert\copyrightnote@ins
          {%
            \reset@font\footnotesize
            \interlinepenalty\interfootnotelinepenalty
            \splittopskip\footnotesep
            \splitmaxdepth\dp\strutbox
            \floatingpenalty\@MM
            \hsize\columnwidth
            \@parboxrestore
            \color@begingroup
            \parindent1em
            \noindent
            \hskip1.8em
            \ignorespaces #1\@finalstrut\strutbox
            \color@endgroup
          }%
      }%
  }
\makeatother

\copyrightnote{Accepted for publication in ICCS 2020 conference proceedings.}
\copyrightnote{The final authenticated publication is available online at \url{https://doi.org/10.1007/978-3-030-50371-0_51}.}

\begin{document}

\maketitle

\begin{abstract}
In this work we present useful techniques and possible enhancements when applying an Algorithmic Differentiation (AD) tool to the linear algebra library Eigen using our in-house AD by overloading (AD-O) tool \dcocpp{} as a case study. After outlining performance and feasibility issues when calculating derivatives for the official Eigen release, we propose \textit{Eigen-AD}, which enables different optimization options for an AD-O tool by providing add-on modules for Eigen. The range of features includes a better handling of expression templates for general performance improvements as well as implementations of symbolically derived expressions for calculating derivatives of certain core operations. The software design allows an AD-O tool to provide specializations to automatically include symbolic operations and thereby keep the look and feel of plain AD by overloading. As a showcase, \dcocpp{} is provided with such a module and its significant performance improvements are validated by benchmarks.

\end{abstract}
  
\section{Introduction}
\label{sec:introduction}
In this work, the \cpp{} linear algebra library Eigen\footnote{\url{http://eigen.tuxfamily.org}} is used as a base software implementing linear algebra operations for which derivatives are to be computed using Algorithmic Differentiation (AD)~\cite{AD1,AD2} by overloading. Derivatives of computer programs can be of interest, e.g.\ when performing uncertainty quantification~\cite{doi:10.1080/10556788.2017.1359267}, sensitivity analysis~\cite{doi:10.1137/S1064827503426723} or shape optimization~\cite{10.1007/978-3-540-74460-3_56}. AD enables the computation of derivatives of the output of such programs with respect to their inputs. This is done using the tangent model in \textit{tangent mode} or the adjoint model in \textit{adjoint mode}, where the latter is also known as adjoint AD (AAD). In AAD, the program is first executed in the \textit{augmented primal run}, where required data for later use is stored. Derivative information is then propagated through the tape in the \textit{adjoint run}. For both, tangent and adjoint, the underlying original code is called the \textit{primal}, and the used floating point data type and its variables are called \textit{passive}. Vice versa, code where derivatives are computed and its respective data type and variables are called \textit{active}.
\par
A wide collection of AD tools can be found on the community website\footnote{\url{http://www.autodiff.org/}}. In general, one can divide the available software into \textit{source transformation} and \textit{operator overloading} tools. While source transformation essentially has the potential to create more efficient code, supporting complex language features like they are available in \cpp{} is connected to higher expense for the tool authors. AD by overloading (AD-O) on the other hand can be applied to arbitrary code as long as operator overloading is supported by the programming language. In terms of AD-O, the recorded data in the augmented primal run is referred to as the \textit{tape}, and creating the tape is called \textit{taping}. The propagation in the adjoint run is known as \textit{interpreting} the tape. 
\par
Applying an AD-O tool to dedicated libraries poses a significant issue, as by principle they require the usage of an extended floating point data type (from now referred to as the \textit{custom AD data type}). This change in data type is often impractical and breaks hand tuned performance gains~\cite{Dunham2017HOA}. Therefore, software combining AD-O and linear algebra has been realized with, e.g.\ Adept~\cite{adept} or the \textit{Stan Math Library}\cite{stanMath}, where the latter makes heavy use of Eigen. Eigen allows the direct utilization of AD-O tools due to its extensive use of \cpp{} templates. At a later point in this paper, concrete implementations and benchmarks for AD-O in Eigen will be presented using \dcocpp{}, which is an AD-O tool actively developed by NAG Ltd.\footnote{\url{https://www.nag.co.uk/}} in collaboration with RWTH Aachen University.
\par
To our best knowledge, there has not been a work focusing on the application of an AD-O tool to Eigen while preserving the philosophy of plain AD-O. The goal is that the AD-O tool user benefits from optimizations without explicitly being aware of them. Swapping the data type of Eigen and using the AD-O tool as usual should be all that is required to compute derivatives. However, several problems concerning performance and feasibility of the derivative computation will arise from this concept. This work proposes approaches and solutions to overcome them.
\par
The next section provides more background on AD-O and also introduces the concept of symbolic derivatives. Section~\ref{sec:eigenad} presents \textit{Eigen-AD} which is a fork of Eigen. It contains several optimizations and improvements for the application of an AD-O tool, which are demonstrated and benchmarked using \dcocpp{} in Section~\ref{sec:bench}. Section~\ref{sec:ConclusionOutlook} summarizes the results and suggests possible future works. 
\par
Note than an extended version of this work exists~\cite{peltzer2019eigenad}; refer to it for further details.

\section{Using AD-O and Symbolic Derivatives}
\label{sec:adVsSym}
Most of the performance improvements presented at a later point in this paper are based on symbolic differentiation (SD), in which derivatives are evaluated analytically at a higher level than with AD. This section demonstrates the differences between evaluating derivatives symbolically and with AD-O by using the matrix-matrix product \(C = AB\) with \(A, B \in \mathbb{R}^{2\times2}\) as an example.
\par
\begin{figure}[t]
 \centering
	\begin{tikzpicture}
	\tikzset{VertexStyle/.style = {shape=rectangle, rounded corners, draw, align=center, top color=white, bottom color=blue!20}}
	\tikzset{EdgeStyle/.style = {-{Latex[scale=1.5]}}}
	\tikzset{LabelStyle/.style= {sloped, above, fill opacity=0, text opacity=1, font=\scriptsize}}
	\Vertex[Math, x=-4.5,y=0]{C_{0,0}}
	\Vertex[Math, x=1.5 ,y=0]{C_{1,0}}
	\Vertex[Math, x=-1.5 ,y=0]{C_{1,1}}
    \Vertex[Math, x=4.5,y=0]{C_{0,1}}
	
	\Vertex[Math, x=-5.25,y=-1]{z_0}
	\Vertex[Math, x=-3.75,y=-1]{z_1}
    \Vertex[Math, x=-2.25,y=-1]{z_2}
    \Vertex[Math, x=-0.75,y=-1]{z_3}
    \Vertex[Math, x=0.75 ,y=-1]{z_4}
	\Vertex[Math, x=2.25 ,y=-1]{z_5}
	\Vertex[Math, x=3.75 ,y=-1]{z_6}
	\Vertex[Math, x=5.25 ,y=-1]{z_7}
	
    \Vertex[Math, x=-5.25,y=-4]{A_{0,0}}
	\Vertex[Math, x=-3.75,y=-4]{A_{0,1}}
	\Vertex[Math, x=-2.25,y=-4]{A_{1,0}}
	\Vertex[Math, x=-0.75,y=-4]{A_{1,1}}
    \Vertex[Math, x=0.75 ,y=-4]{B_{0,0}}
	\Vertex[Math, x=2.25 ,y=-4]{B_{0,1}}
	\Vertex[Math, x=3.75 ,y=-4]{B_{1,0}}
	\Vertex[Math, x=5.25 ,y=-4]{B_{1,1}}
	
	\Edge[label=$B_{0,0}$, labelstyle={pos=0.5}](A_{0,0})(z_0)
	\Edge[label=$A_{0,0}$, labelstyle={pos=0.65}](B_{0,0})(z_0)
	
    \Edge[label=$B_{1,0}$, labelstyle={pos=0.5}](A_{0,1})(z_1)
	\Edge[label=$A_{0,1}$, labelstyle={pos=0.06}](B_{1,0})(z_1)
	
	\Edge[label=$B_{0,1}$, labelstyle={pos=0.07}](A_{0,0})(z_5)
	\Edge[label=$A_{0,0}$, labelstyle={pos=0.35}](B_{0,1})(z_5)
	
	\Edge[label=$B_{1,1}$, labelstyle={pos=0.06}](A_{0,1})(z_7)
	\Edge[label=$A_{0,1}$, labelstyle={pos=0.5}](B_{1,1})(z_7)
	
	\Edge[label=$B_{0,0}$, labelstyle={pos=0.75}](A_{1,0})(z_4)
	\Edge[label=$A_{1,0}$, labelstyle={pos=0.6}](B_{0,0})(z_4)
	
	\Edge[label=$B_{1,0}$, labelstyle={pos=0.17}](A_{1,1})(z_6)
	\Edge[label=$A_{1,1}$, labelstyle={pos=0.5}](B_{1,0})(z_6)
	
	\Edge[label=$B_{0,1}$, labelstyle={pos=0.65}](A_{1,0})(z_2)
	\Edge[label=$A_{1,0}$, labelstyle={pos=0.83}](B_{0,1})(z_2)
	
	\Edge[label=$B_{1,1}$, labelstyle={pos=0.1}](A_{1,1})(z_3)
	\Edge[label=$A_{1,1}$, labelstyle={pos=0.07}](B_{1,1})(z_3)
	
	\Edge[label=$1$, labelstyle={pos=0.5}](z_0)(C_{0,0})
	\Edge[label=$1$, labelstyle={pos=0.5}](z_1)(C_{0,0})
	
	\Edge[label=$1$, labelstyle={pos=0.5}](z_5)(C_{0,1})
	\Edge[label=$1$, labelstyle={pos=0.5}](z_7)(C_{0,1})
	
	\Edge[label=$1$, labelstyle={pos=0.5}](z_4)(C_{1,0})
	\Edge[label=$1$, labelstyle={pos=0.5}](z_6)(C_{1,0})
	
    \Edge[label=$1$, labelstyle={pos=0.5}](z_2)(C_{1,1})
	\Edge[label=$1$, labelstyle={pos=0.5}](z_3)(C_{1,1})

\end{tikzpicture}
	\caption{Computational graph for the \(2 \times 2\) matrix product kernel displayed in Listing~\ref{lst:matmulKernel}: Vertices represent variables accessed in the augmented primal run, edges define computation operands and edge weights are the respective partial derivatives. Temporaries \(z\) storing the results of the \lstinline{*}{} operator in line 6 may be optimized away by an AD-O tool when taping.}
	\label{fig:compGraph}
\end{figure}
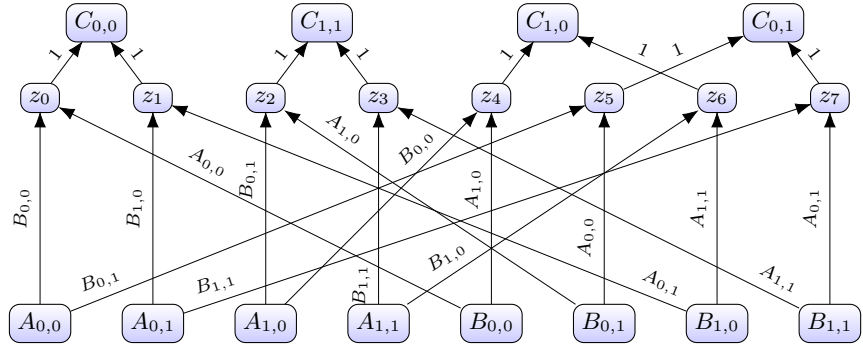

Let this specific product kernel be implemented using Eigen as follows:
\begin{lstlisting}[caption=\(2 \times 2\) matrix-matrix multiplication kernel.,,captionpos=b,label=lst:matmulKernel]
template<typename T> 
void matmul(const Matrix<T,2,2>& A, const Matrix<T,2,2>& B, Matrix<T,2,2>& C)  {
  for(int i=0; i<2; i++) 
    for(int j=0; j<2; j++) 
      for(int k=0; k<2; k++) 
        C(i,j) += A(i,k)*B(k,j);
}
\end{lstlisting}
The primal code is called using the passive data type \lstinline[keywordstyle={}]{double} 
as template argument \lstinline{T}{}. For an active evaluation, the function must be called using the custom AD data type of an AD-O tool as \lstinline{T}{}. As mentioned in the previous section, the AD-O tool first performs the augmented primal run when in adjoint mode. Both, the \lstinline{+=}{} and the \lstinline{*}{} operators in line 6, are overloaded by the tool and act as the entry points for taping. The tape is an implementation dependent representation of the \textit{computational graph} of the program, which contains all performed computations and their corresponding partial derivatives. Fig.~\ref{fig:compGraph} displays the computational graph of the matrix-matrix multiplication kernel in Listing~\ref{lst:matmulKernel}. Vertices represent variables accessed in the augmented primal run, including temporary instantiations from the \lstinline{*}{} operator in line 6 (denoted as $z$). The edge weights are the partial derivatives of the respective computations. In the adjoint run, the graph is traversed in reverse order, propagating the adjoint value of the output towards the inputs. This is done by multiplying subsequent edge weights and adding parallel edge weights. Effectively, the loops of Listing~\ref{lst:matmulKernel} are executed in reverse order; derivatives are computed on \emph{scalar level}.
\par
In contrast to the differentiation of all occurring scalar computations, it may also be possible to rewrite the derivative using matrix expressions so that derivatives are computed on \emph{matrix level}. Staying with the example above, the adjoint propagation on the computational graph in Fig.~\ref{fig:compGraph} can be written as follows:
\\
\begin{minipage}{.5\textwidth}
\begin{align}
  \nonumber \bar{A}_{0,0} &= \bar{C}_{0,0} B_{0,0} + \bar{C}_{0,1} B_{0,1}\\
  \nonumber \bar{A}_{0,1} &= \bar{C}_{0,0} B_{1,0} + \bar{C}_{0,1} B_{1,1}\\
  \nonumber \bar{A}_{1,0} &= \bar{C}_{1,0} B_{0,0} + \bar{C}_{1,1} B_{0,1}\\
  \nonumber \bar{A}_{1,1} &= \bar{C}_{1,0} B_{1,0} + \bar{C}_{1,1} B_{1,1}\\
  \nonumber\\
  \Rightarrow \bar{A} &= \bar{C} B^T \label{eq:sym1}
\end{align}
\end{minipage}
\begin{minipage}{.5\textwidth}
\begin{align}
  \nonumber \bar{B}_{0,0} &= \bar{C}_{0,0} A_{0,0} + \bar{C}_{1,0} A_{1,0}\\
  \nonumber \bar{B}_{0,1} &= \bar{C}_{0,1} A_{0,0} + \bar{C}_{1,1} A_{1,0}\\
  \nonumber \bar{B}_{1,0} &= \bar{C}_{0,0} A_{0,1} + \bar{C}_{1,0} A_{1,1}\\
  \nonumber \bar{B}_{1,1} &= \bar{C}_{0,1} A_{0,1} + \bar{C}_{1,1} A_{1,1}\\
  \nonumber\\
  \Rightarrow \bar{B} &= A^T \bar{C} \label{eq:sym2}
\end{align}
\end{minipage}
\\[0.5cm]
Adjoint values are denoted with a bar. Equations~(\ref{eq:sym1}) and (\ref{eq:sym2}) compute the adjoints of the input data using matrix-matrix multiplications. Using these equations, it is not necessary to tape any computations in Listing~\ref{lst:matmulKernel}. Instead, the adjoints can directly be computed in the adjoint run on matrix level as long as the values of the input data are available.

\section{Eigen-AD}
\label{sec:eigenad}
Applying an AD-O tool to Eigen will lead to severe limitations sooner or later. Eigen comes with optimized kernels, e.g.\ for 8-byte double precision data. Traditionally, custom AD data types are larger and these optimizations do not work anymore. Regarding AAD application, the complexity of many frequently used linear algebra operations scales cubically with the input dimension. This is the case for, e.g.\ matrix decompositions or matrix products. Since the memory required by the tape scales roughly linearly with the number of operations required by an algorithm, the tape size can quickly exceed the amount of available RAM and therefore makes an evaluation of the derivatives not feasible at all. 
\par
\begin{figure}[t]
	\centering
		\includegraphics[width=1\textwidth]{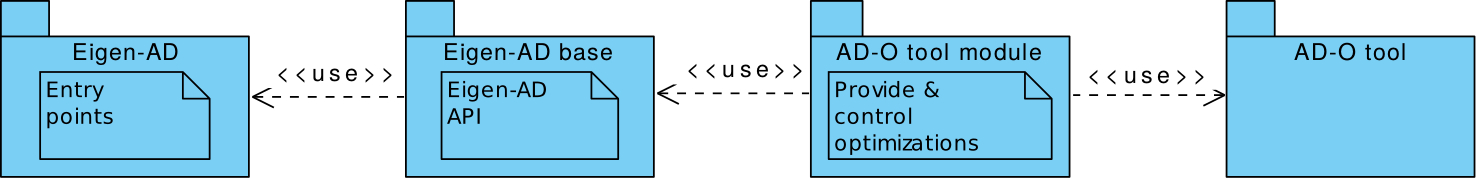}
                \caption[caption]{Eigen-AD package architecture: Authors can implement an AD-O tool module for their AD-O tool.}
		\label{fig:classDiagram4}
\end{figure}

To overcome these issues, the Eigen source code has been adjusted and extended to help optimize the application of AD-O tools. The resulting software has been named \emph{Eigen-AD}. All source code changes are generically written and do not modify the original Eigen API, but provide entry points which can be used by additional modules. Based on that, we have added a generic Eigen-AD base module which provides a clean interface for developers to control and implement optimized operations in their tool specific AD-O tool module. Refer to Fig.~\ref{fig:classDiagram4} for the package architecture.
\par
The existing Eigen test system has been extended so that every Eigen test can also be performed for an AD-O tool's tangent and adjoint data types. Compiling and running the tests successfully ensures compatibility of the AD-O tool with all of the tested Eigen functions. The philosophy is that an AD-O tool is able to determine derivatives of all Eigen operations algorithmically, while selected operations are provided with optimized computations for their derivatives. Another aim is to keep the look and feel of AD-O, i.e.\ optimizations and improvements shall not require a separate interface but be used automatically whenever the AD-O tool is applied.
\par
The next sections present optimization approaches realized in Eigen-AD.

\subsection{Nesting Expression Templates}
\label{ssec:nestingExpTem}
The concept of expression templates was originally proposed to eliminate temporaries when evaluating vector and matrix expressions and to be able to pass algebraic expressions as arguments to functions~\cite{et1}. The first aspect, also known as \textit{lazy evaluation}, has been complemented by the concept of \textit{Smart Expression Templates}~\cite{et2} which both are implemented in Eigen.
\par
In the context of AD, using expression templates is especially relevant, since every temporary contributes to the computational graph as it was also demonstrated in Fig.~\ref{fig:compGraph} in Section~\ref{sec:adVsSym}. In the AAD case, the computational graph needs to be stored in memory and is then traversed in the reverse run, increasing memory and run time requirements with each additional temporary. This can be avoided by constructing expression templates for the right hand side of the assignment and evaluate them altogether. Therefore, some AD-O tools also implement an expression template mechanism, e.g.\ \dcocpp{} or Adept ~\cite{dcocppUserguide,adept}.
\par
When applying such an AD-O tool to Eigen, both expression template engines are nested, where the AD-O tool layer is accessed by the scalar operations of Eigen. This is not an intended use case for Eigen, and therefore Eigen is not aware that it may receive template expressions. The returned template expression is then implicitly casted back to the custom AD data type, resulting in a temporary which must be considered for the derivative evaluation. This destroys the gains originally made by using expression templates in the AD-O tool.
\par
As an example, consider the unary minus operator, implemented in Eigen as a functor named \lstinline{scalar_opposite_op}{}. Its class template parameter \lstinline{Scalar}{} corresponds to the custom AD data type and it is also used in the parenthesis operator as the in- and output types. An assignment of the form \(A=-B\), where \(A\) and \(B\) are Eigen \(1 \times 1\) matrices containing a single scalar of the adjoint data type, will result in an additional vertex in the computational graph, analogous to the temporaries \textit{z} of the computational graph in Fig.~\ref{fig:compGraph}.
\par
When looking at the way the Eigen functors are used, it is not necessary to explicitly prescribe what types they return. Due to Eigen's generic design and as long as the occurring types are compatible -- meaning the required casts/specializations/overloads are available -- there is no need to force the scalar type at this level. This is a fitting case to use the \cpp{}-14 feature of \textit{auto return type deduction} which allows a function to deduce the return type from the operand of its return statement. Therefore, replacing the return type of the functors with the \textit{auto} keyword allows the passing of expression types from the AD-O tool to the Eigen internals. Besides that, it must be ensured that the functors allow arbitrary input types, as they can now be called with expression types as parameters as well.
\par
Evaluating the modified \lstinline{scalar_opposite_op}{} functor will avoid the additional vertex in the computational graph. This optimization can be applied to all Eigen scalar functors and also to several Eigen math functions like \lstinline{sin}{} or \lstinline{exp}{} which are supported by the AD-O tool's expression templates.

\subsection{Symbolic Derivatives}
\label{ssec:symImpl}
As introduced in Section~\ref{sec:adVsSym}, mathematical insight can be exploited to evaluate a derivative symbolically. Such an evaluation can be superior to the AD-O solution in terms of performance, run time-wise and also memory-wise in the adjoint case. This observation motivates the inclusion of symbolic derivatives for certain linear algebra routines, yielding a \textit{hybrid} implementation~\cite{thesisLotz}.
\par
The Eigen-AD base module provides an interface for AD-O tool developers to implement symbolic derivatives. At the moment, entry points for products as well as for any computation concerning a dense solver are supported. Refer to the Eigen-AD base module technical guide for further information. In the next sections, equations for symbolic adjoints of selected operations are introduced.

\subsubsection{SD Dense System Solver}
\label{sssec:symbolicSolvers}
Consider the system of linear equations:
\begin{align}
A \textbf{x} = \textbf{b} \label{eq:linearEquationSystem}
\end{align}
where \( A \in \mathbb{R}^{n \times n} \) is the system matrix, \( \textbf{b} \in \mathbb{R}^n \) is the right hand side vector and \( \textbf{x} \in \mathbb{R}^n \) is the solution vector. There exists a wide variety of approaches to solve the problem shown in Equation~(\ref{eq:linearEquationSystem}) which make use of decomposing the matrix \( A \) into a product of other matrices, e.g. the \textit{LU decomposition}. Eigen offers one dense solver class for each decomposition type.
\par
AAD for the solution of a system of linear equation includes the taping and the interpretation of the decomposition, which yields a run time and memory overhead of \( \mathcal O (n^3) \). However, when evaluating the adjoints symbolically using Equations~(\ref{eq:adjointsB})-(\ref{eq:adjointsA}) as presented in~\cite{collectedMatrixDerivative}, the decomposition is completely excluded from taping and interpreting.
\begin{align}
A^T \cdot \bar{\textbf{b}} & = \bar{\textbf{x}} \label{eq:adjointsB} \\
\bar{A} & = - \bar{\textbf{b}} \cdot \textbf{x}^T \label{eq:adjointsA}
\end{align}
As it can be seen, the adjoint values of the right hand side vector \(\textbf{b}\) can be determined by solving an additional linear system. By saving the computed decomposition of \(A\) in the augmented primal run, it can then be reused in the adjoint run. This reduces the run time and memory overhead for differentiating to \( \mathcal O (n^2) \)~\cite{NL2012}.

\subsubsection{Symbolic Inverse}
Inverting a matrix, i.e.\ computing
\begin{align}
C = A^{-1} \label{eq:inverse}
\end{align}
is implemented in Eigen as a member function of a dense solver. Corresponding adjoints can be computed using Equation~(\ref{eq:symInv})~\cite{collectedMatrixDerivative}.
\par
\begin{align}
\bar{A} & = - C^T \bar{C} C^T \label{eq:symInv}
\end{align}
Compared to AAD, the memory overhead is reduced to \( \mathcal O (n^2) \); however, the adjoint run still has a run time overhead of \( \mathcal O (n^3) \) due to the matrix multiplications.

\subsubsection{Symbolic Log-Abs-Determinant}
Another member function of the dense solvers is the computation of \(x \in \mathbb{R}\) using the log-abs-determinant of a matrix \(A \in \mathbb{R}^{n\times n}\) as shown in Equation~(\ref{eq:logAbsDet}). Such a computation is relevant for, e.g.\ computing the log-likelihood of a multivariate normal distribution.
\begin{align}
x &= \log \left| \det(A) \right| \label{eq:logAbsDet}\\
\bar{A} &= A^{-T} \bar{x} \label{eq:symLogAbsDet}
\end{align}
Equation~(\ref{eq:logAbsDet}) is implemented in Eigen for the QR dense solvers, and adjoints can be computed according to Equation~(\ref{eq:symLogAbsDet})~\cite{IMM2012-03274}. The inverse can be computed by reusing the decomposition which was kept in memory for the adjoint run. While the run time overhead is still \( \mathcal O (n^3) \), the symbolic implementation improves the memory overhead to \( \mathcal O (n^2) \).

\subsubsection{Symbolic Matrix-Matrix Product}
\label{sssec:symMatMat}
For \(A \in \mathbb{R}^{n\times m}, B \in \mathbb{R}^{m\times p}, C \in \mathbb{R}^{n\times p}\), the adjoints of the matrix-matrix product in Equation~(\ref{eq:symProd}) can be computed using Equations~(\ref{eq:symProdA})-(\ref{eq:symProdB}) according to~\cite{collectedMatrixDerivative}.
\begin{align}
C & = A B \label{eq:symProd}\\
\bar{A} & = \bar{C} B^T \label{eq:symProdA}\\
\bar{B} & = A^T \bar{C} \label{eq:symProdB}
\end{align}
Note that this matches the results derived in Section~\ref{sec:adVsSym} for the \(2 \times 2\) matrix-matrix product. While differentiating the matrix-matrix product with AAD has a run time and memory complexity of \( \mathcal O (nmp) \), utilizing the symbolic evaluation reduces the memory overhead to \( \mathcal O (nm+mp) \). The input matrices \(A\) and \(B\) must be saved in the augmented primal run and then be multiplied with the adjoint values of the output according to Equations~(\ref{eq:symProdA})-(\ref{eq:symProdB}) in the adjoint run. Note that the run time complexity can not be improved using the symbolic evaluation.

\section{Benchmarks}
\label{sec:bench}
As mentioned in the beginning of this paper, an Eigen-AD tool module has been implemented for \dcocpp{}. In order to verify the implementation, extensive measurements were made. They were performed using a single thread on an i7-6700K CPU running at \SI{4}{\giga\hertz} with AVX2 enabled and \SI{64}{\giga\byte} RAM available for the tape recording, using the g++ 7.4 compiler on Ubuntu 18.04. The respective linear algebra operations were performed for increasing matrix size \(n\) using dynamic-sized quadratic matrices \(\mathbb{R}^{n\times n}\) and one evaluation of the first-order adjoint model was computed with all output adjoints set to 1. The \lstinline{inverse()}{} results shown here use the underlying \lstinline{PartialPivLU}{}, the \lstinline{logAbsDeterminant()}{} function the \lstinline{FullPivHouseholderQR}{} solver. From now on, the \dcocpp{} Eigen module is referred to as \dcocppeigen{}, and computations which are not using symbolic implementations but only plain overloading are denoted as algorithmic or as AD.

\subsection{\dcocppeigen{} benchmarks}
In this section, the theoretical considerations from Section~\ref{ssec:symImpl} are validated with benchmarks. To emphasize the improvements, reference measurements for the corresponding algorithmic computations without the \dcocppeigen{} module are given where appropriate.
  
\subsubsection{Memory consumption}
\begin{figure}[t]
\centering
\subcaptionbox{\lstinline{solve(b)}{} of selected solvers\label{fig:benchTapeSize1}}%
  [.49\linewidth]{\includegraphics[width=0.45\textwidth]{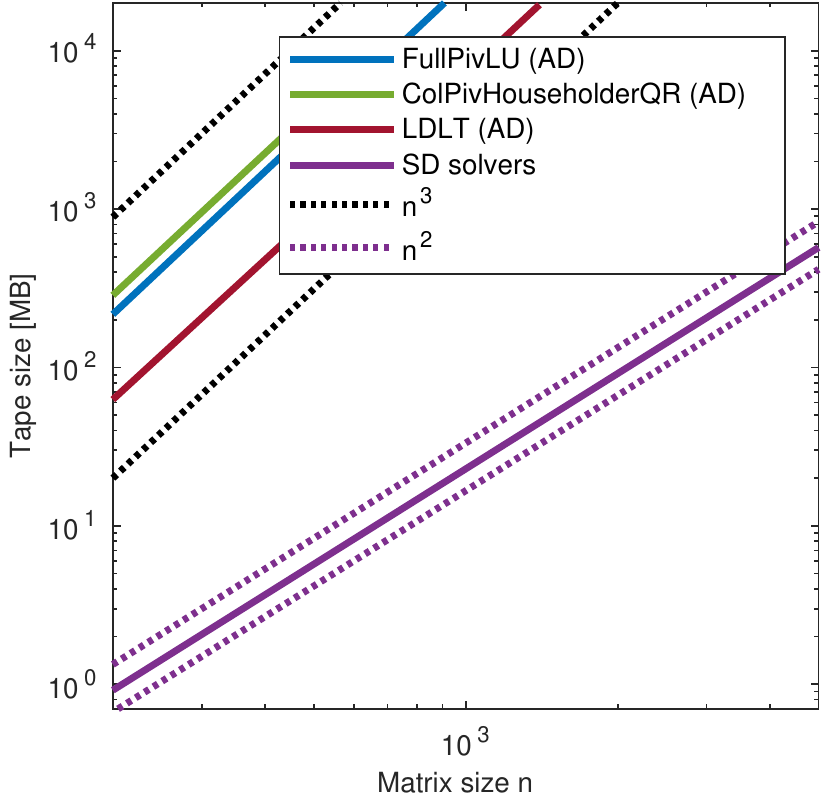}}
\subcaptionbox{Other symbolic operations\label{fig:benchTapeSize2}}
  [.49\linewidth]{\includegraphics[width=0.45\textwidth]{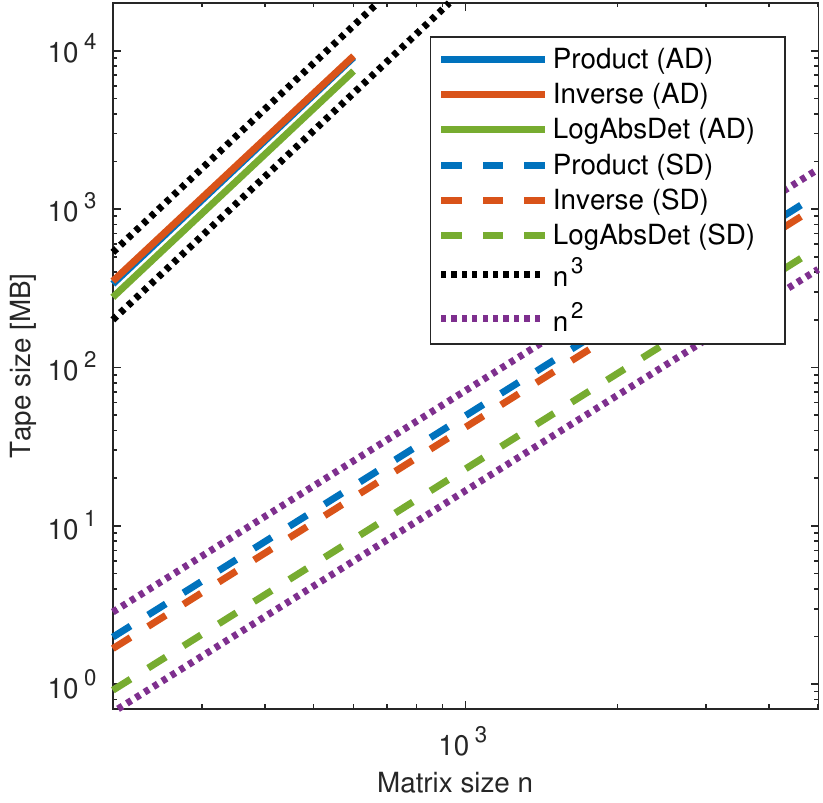}}
    \caption{Tape size comparison between AD and SD solvers: All new symbolic implementations have a memory complexity of \( \mathcal O (n^2) \) compared to \( \mathcal O (n^3) \) of the algorithmic versions.}
  \label{fig:benchTapeSize}
\end{figure}

All symbolic evaluations introduced in Section~\ref{ssec:symImpl} lower the memory overhead introduced by an AD-O tool to \( \mathcal O (n^2) \). In order to visualize this effect, the tape size of \dcocpp{} has been measured for the algorithmic and for the symbolic implementations and is displayed in Fig.~\ref{fig:benchTapeSize}. For clarity reasons, only selected dense solvers are shown in Fig.~\ref{fig:benchTapeSize1}, but similar patterns were measured for the other solvers as well. The symbolic implementations keep a complete primal solver in memory which is accounted with a \(n\times n\) matrix on the tape. Since the symbolic \lstinline{logAbsDeterminant()}{} function does not require any additional data, it has the same memory usage as its corresponding solver. The symbolic \lstinline{inverse()}{} function additionally saves the transposed input matrix, the symbolic matrix-matrix product stores the two input matrices.\par
As it was expected, all new implementations have a memory complexity of \( \mathcal O (n^2) \), while the algorithmic versions display a cubic behaviour and quickly exceed the amount of available RAM.

\subsubsection{Run time analysis}
\begin{figure}[t]
\centering
\subcaptionbox{\lstinline{solve(b)}{} of selected solvers\label{fig:benchRuntime1}}%
  [.49\linewidth]{\includegraphics[width=0.45\textwidth]{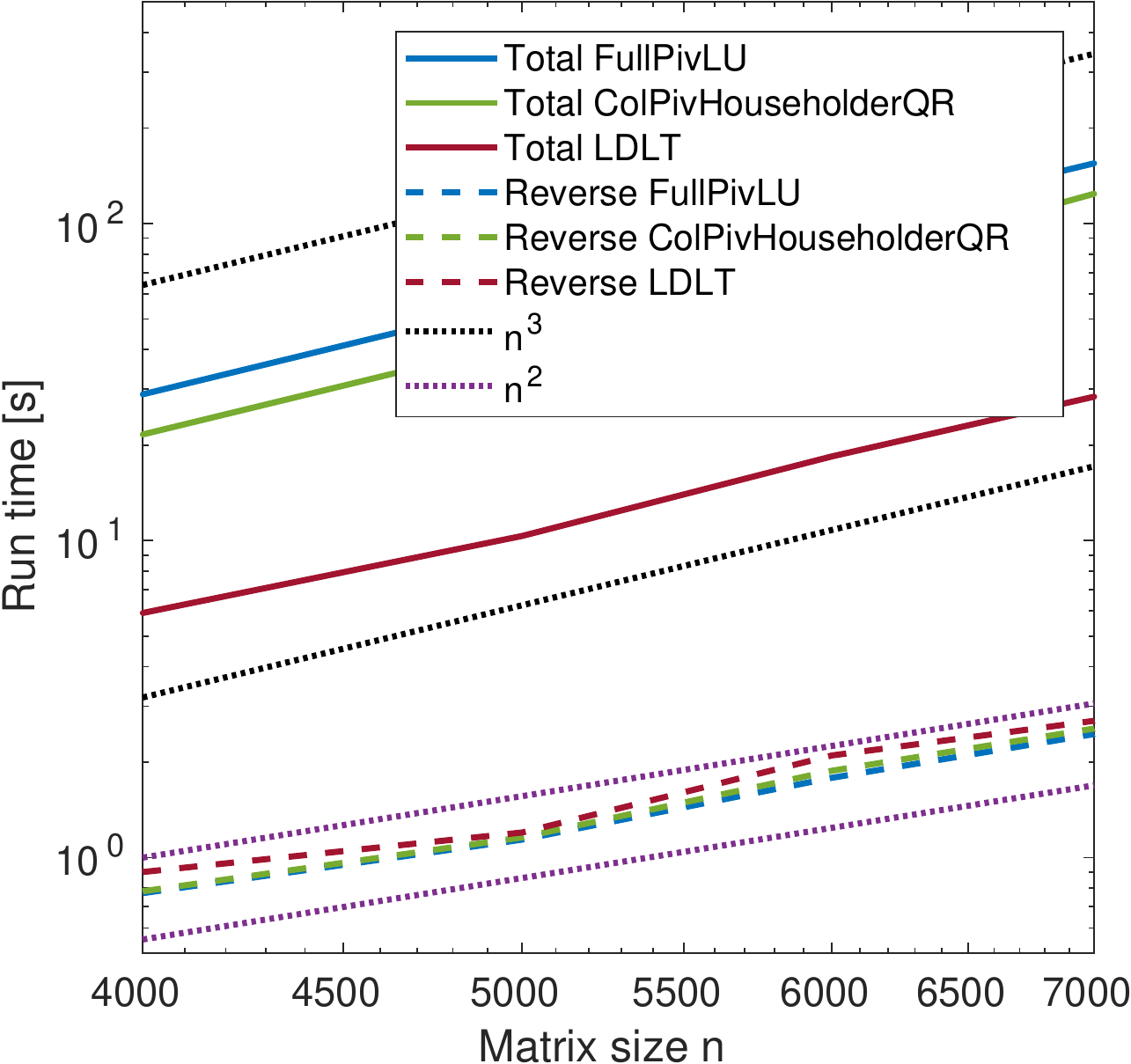}}
\subcaptionbox{Other symbolic operations\label{fig:benchRuntime2}}
  [.49\linewidth]{\includegraphics[width=0.45\textwidth]{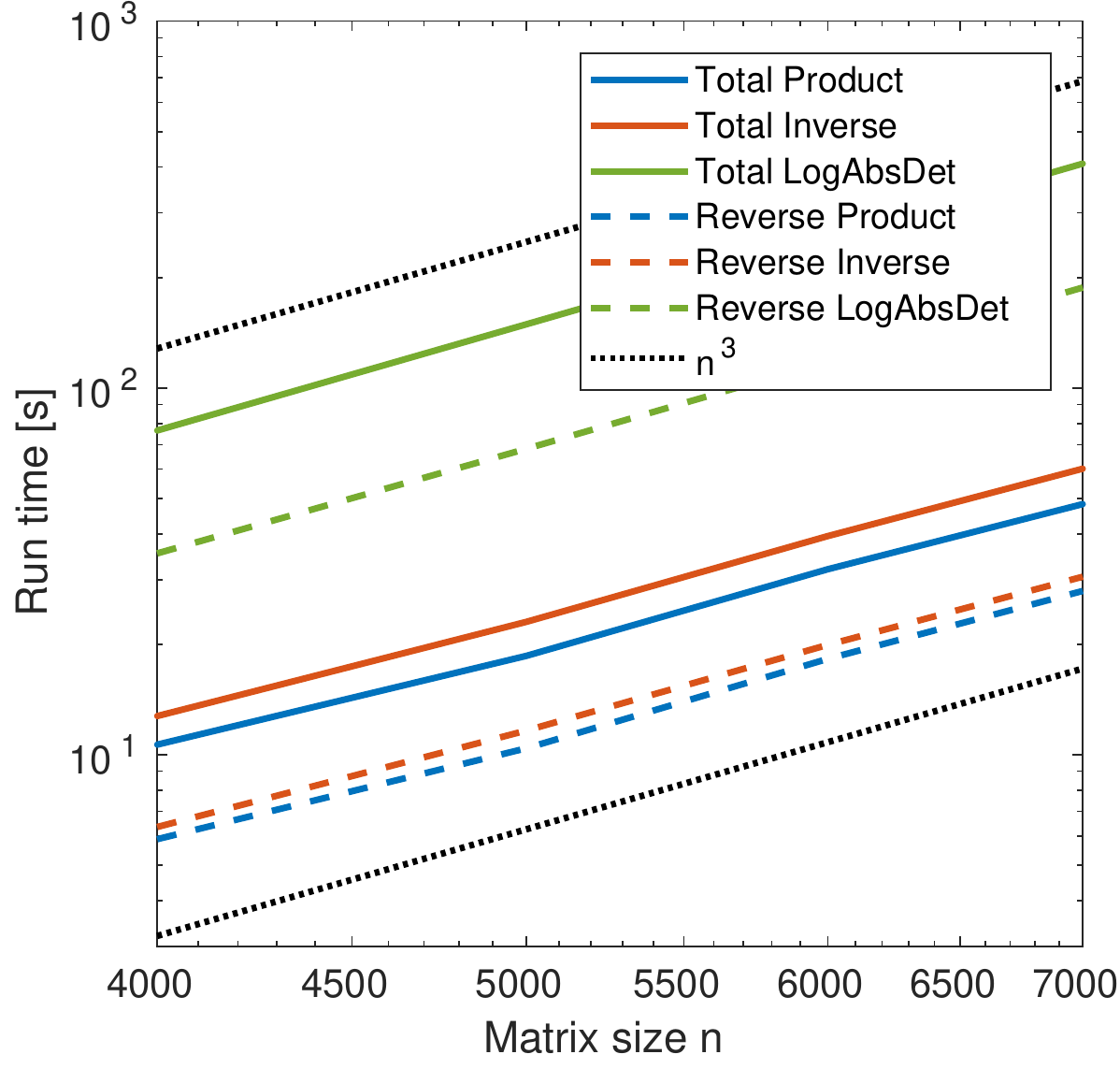}}
    \caption{Total run times and run times of the adjoint run of the new symbolic operations: As described in Section~\ref{sssec:symbolicSolvers}, the symbolic \lstinline{solve(b)}{} function improves the run time complexity of the adjoint run to \( \mathcal O (n^2) \) and effectively cancels the AD overhead compared to the \( \mathcal O (n^3) \) primal run time complexity. The other symbolic operations still have the same run time complexity of the adjoint run as the primal code, but are noticeably faster.}
  \label{fig:benchRuntime}
\end{figure}
Fig.~\ref{fig:benchRuntime} visualizes the run time measurements for the symbolic operations, split into total execution time and run time of the adjoint section. As stated in Section~\ref{sssec:symbolicSolvers}, solving a system of linear equations reduces the adjoint run time to \( \mathcal O (n^2) \), which is confirmed by the measured run times in Fig.~\ref{fig:benchRuntime1}. All other symbolic evaluations do not lower the complexity, since a matrix-matrix product or an inverse must be computed in the adjoint run. However, as it can be inferred from the gap between total and adjoint run times in Fig.~\ref{fig:benchRuntime2}, the overhead introduced by the adjoint run is rather moderate.

\subsubsection{Comparison to primal operations}
\begin{figure}[t]
\centering
\subcaptionbox{\lstinline{solve(b)}{}\label{fig:benchFactors1}}%
  [.49\linewidth]{\includegraphics[width=0.45\textwidth]{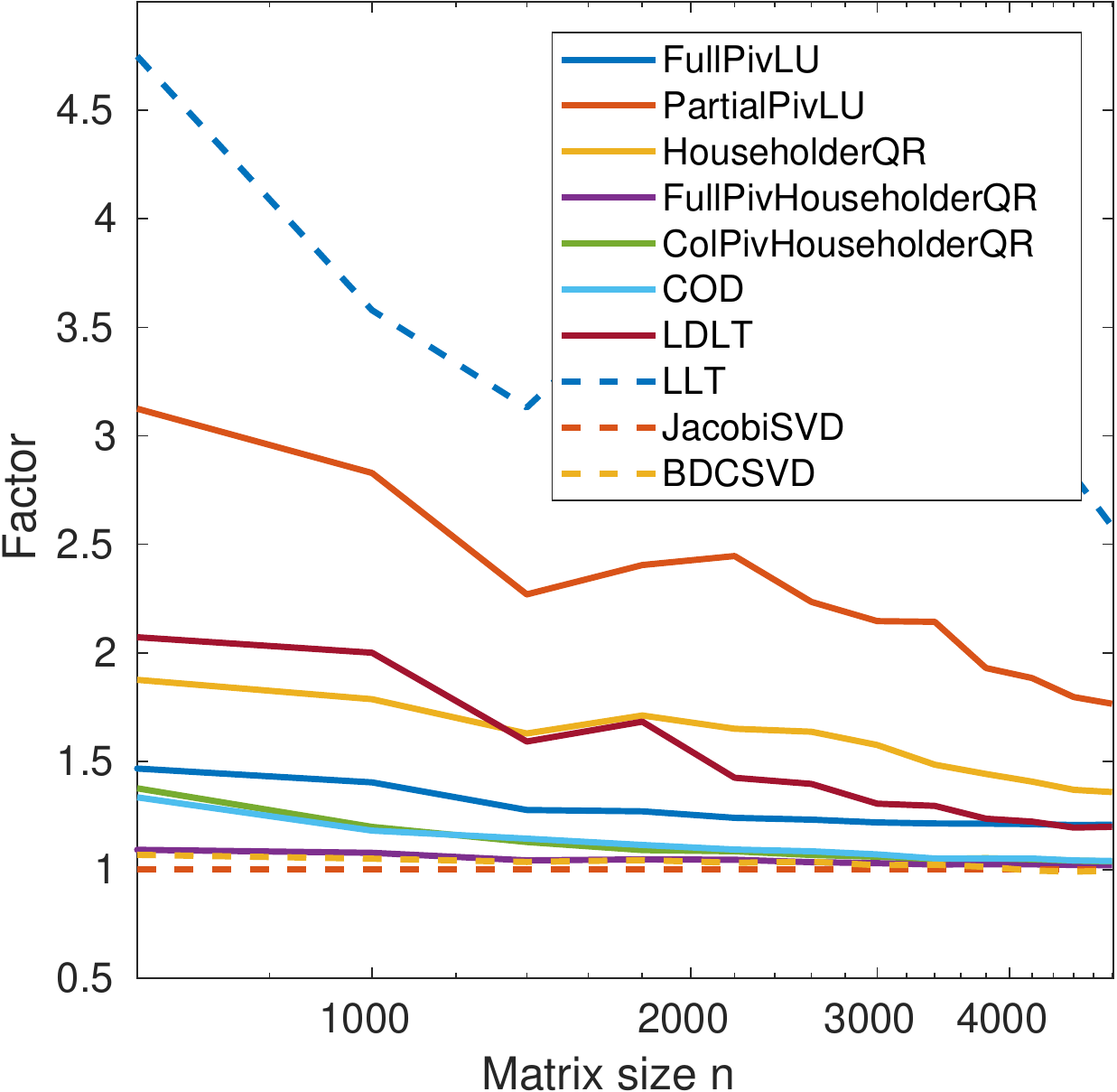}}
\subcaptionbox{Other symbolic operations\label{fig:benchFactors2}}
  [.49\linewidth]{\includegraphics[width=0.45\textwidth]{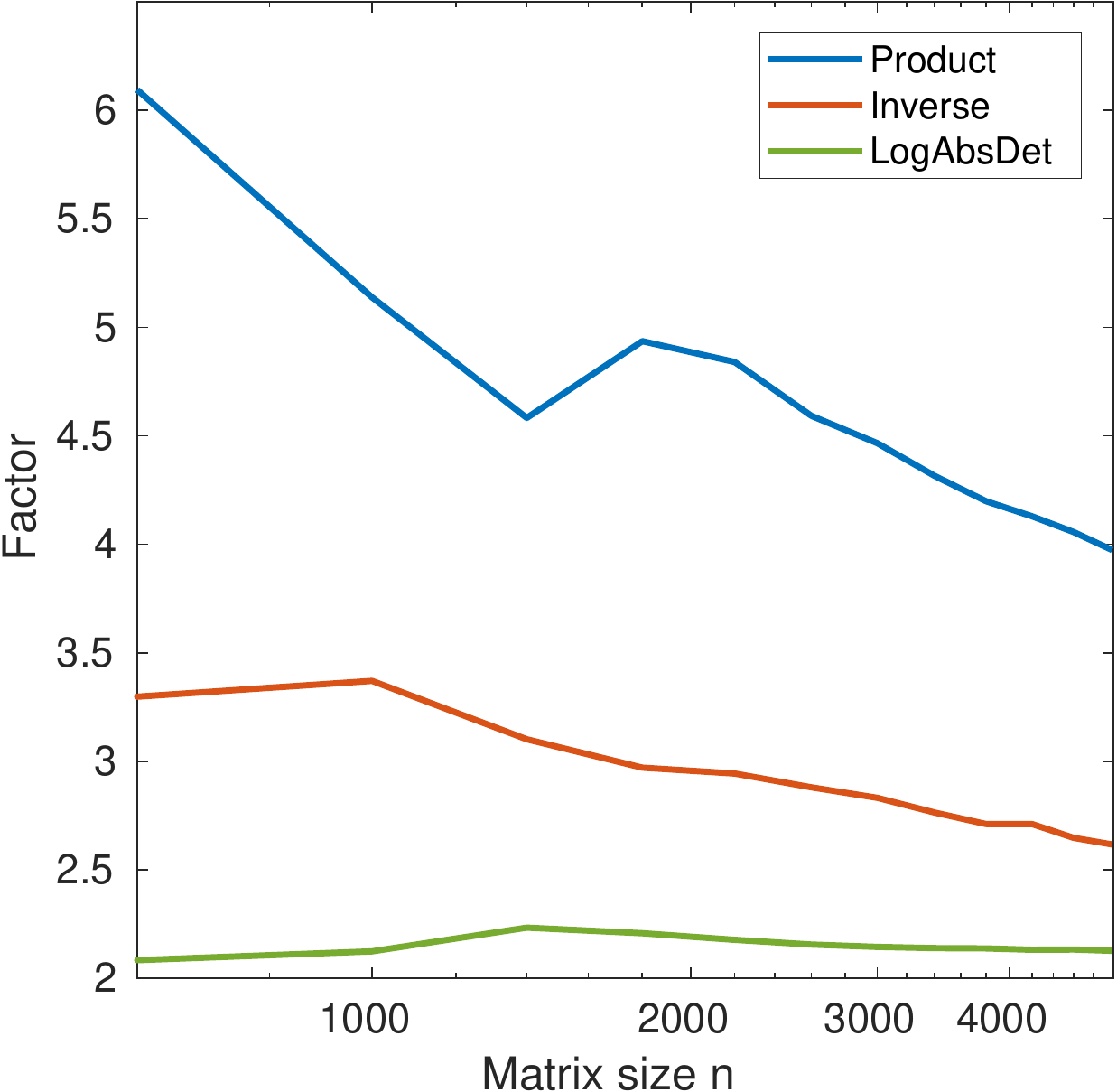}}
    \caption{Run time relation symbolic to primal operations: The run time overhead for the symbolic \lstinline{solve(b)}{} implementation vanishes for larger matrices as all factors converge towards 1. For the other symbolic operations, the overhead does not vanish but still appealing factors are achieved.}
  \label{fig:benchFactors}
\end{figure}

To put the symbolic run times into perspective, the primal run times have been recorded as well. Comparing them both by computing the factor between the respective run times is a good measure to assess the performance of the derivative computation. The results are displayed in Fig.~\ref{fig:benchFactors}.
\par
In contrast to the previous run time analysis, we now compare to the primal code which is highly optimized. Beside the overhead introduced by the AD adjoint section, additional copy instructions are performed in the augmented primal run by the symbolic operations. Due to the convenient fact that the symbolic \lstinline{solve(b)}{} evaluation reduces the run time overhead to \( \mathcal O (n^2) \), all solvers will converge towards a factor of 1 with increasing matrix size, since the ratio is dominated by the \( \mathcal O (n^3) \) primal code. However, as it can be seen in Fig.~\ref{fig:benchFactors1}, the conversion rate depends on the specific solver. For the other symbolic operations, the run time complexity of the adjoint run can not be improved, which makes a factor of 1 impossible. Instead, the factor depends on the additional computations performed in the adjoint run. For the presented symbolic evaluations, an obtainable factor between 2 and 3 is reasonable. Fig.~\ref{fig:benchFactors2} shows a corresponding convergence pattern.
\par
Generically speaking, for very fast primal operations -- like the optimized Cholesky \lstinline{LLT}{} solver or the matrix-matrix product -- it is hard to achieve good factors for smaller input dimensions due to the additional copy overhead introduced by the symbolic implementation. In contrast to that, operations with higher computational costs -- like the \lstinline{JacobiSVD}{} decomposition or the \lstinline{logAbsDet()}{} function -- are dominated by the primal code run time-wise and no significant overhead from the AAD code can be measured even for small matrices.

\subsection{Comparison to other AD-O tools}
To put the above given measurements into perspective, it is reasonable to compare them to results from other AD-O tools. The following tools were considered:
\begin{itemize}
        \item Adept\cite{adept}
        \item ADOL-C\cite{Walther2009GettingSW}
        \item CoDiPack\cite{sagebaum2017high}
		\item FADBAD++ 2.0\footnote{\url{http://www.fadbad.com/fadbad.html}}
        \item Stan Math Library~\cite{stanMath}
\end{itemize}
All tools evaluate a matrix-matrix product of two randomly filled \(\mathbb{R}^{n\times n}\) matrices. One evaluation of the first order adjoint model is performed using plain AD-O of the Eigen library or using the tool's special API if available. Since Stan only provides a \lstinline{grad()}{} function, its benchmark was modified to compute the scalar value \lstinline{z = (A*B).sum()}{} and the corresponding gradient of \textit{z}. All of the following results were produced using Eigen-AD. However, internal benchmarks have not shown a considerable difference to the standard Eigen version.
\par
It must be said that the shown run times do not imply the feasibility of the tools in general, since they are all designed with different use cases and restrictions in mind. They were utilized to our best knowledge, but no tool specific experts were involved in these measurements. While \dcocppeigen{} provides its best performance with this setup, we believe that other tools can be optimized by their developers to get similar results. Therefore, the given results only represent the current situation and are likely to change in the future.
\par
\begin{figure}[t]
\centering
\includegraphics[width=0.95\textwidth]{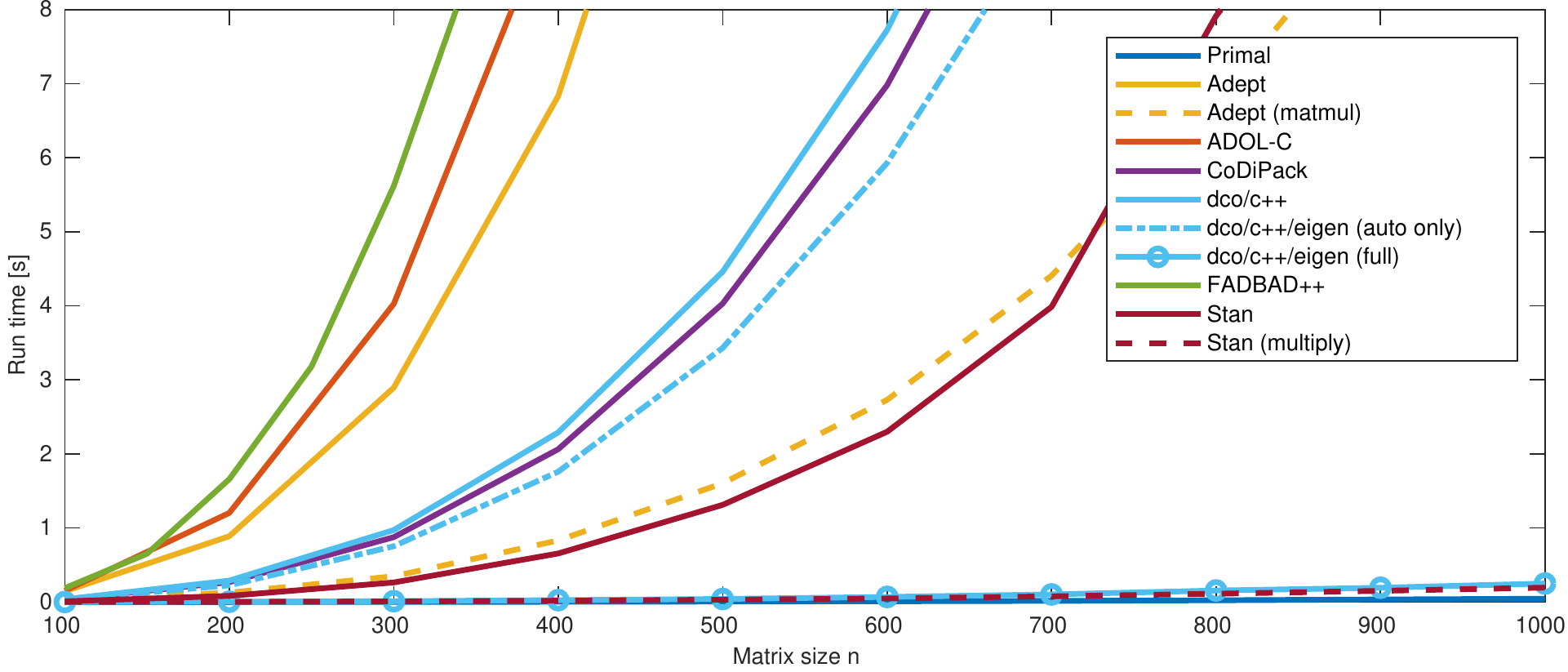}
\caption{Run time comparison of different AD-O tools for the matrix-matrix product: Besides plain algorithmic overloading versions, Adept and Stan also offer optimized functions via their API (denoted by parenthesis).}
\label{fig:benchTools_product}
\end{figure}

The measured run times are displayed in Fig.~\ref{fig:benchTools_product}. Note that the notion \dcocpp{} refers to plain overloading, and the remark \textit{auto only} describes the usage of the \dcocppeigen{} module without any symbolic implementations, i.e.\ only with the optimization from Section \ref{ssec:nestingExpTem} in place. In contrast to that, \textit{full} names the default behaviour when using the module, with the auto return type deduction and symbolic implementations enabled.
\par
All non-specialized tools show the same computational complexity. Differences are non-negligible, though. The feasibility of the auto return type deduction of \dcocppeigen{} introduced in Section~\ref{ssec:nestingExpTem} can be observed, since the smaller amount of temporaries speeds up the computation. In contrast to the other general purpose AD-O tools, Adept also allows the computation of a matrix-matrix product using the \lstinline{matmul}{} function from its API. In this case, no Eigen is used but instead the storage types defined by Adept. As it can be expected, this specially designed feature from Adept is faster than the general AD-O tool approach. This also applies to Stan, although in this case it really is plain AD-O using Eigen storage types. Stan specializes a few Eigen functions such that it internally evaluates optimized matrix-vector products for matrix-matrix products.
\par
\begin{wrapfigure}[24]{r}{0.4\textwidth}
\centering
\includegraphics[width=0.4\textwidth]{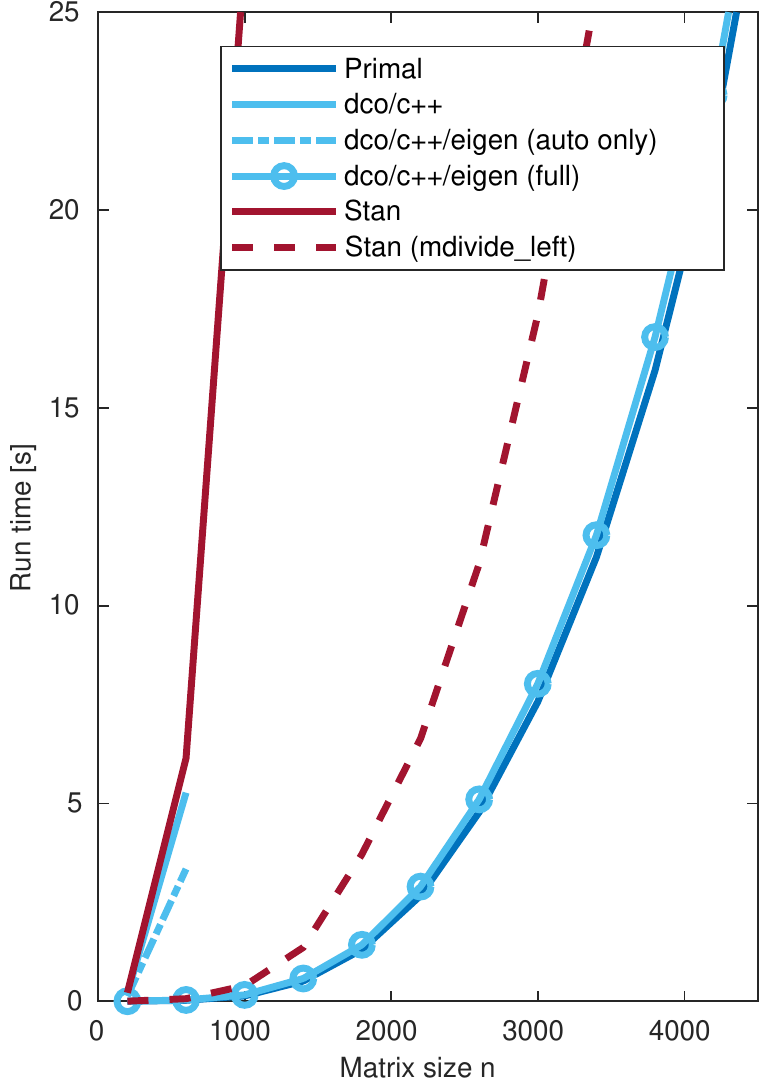}
\caption[]{\label{fig:tools_solver}Algorithmic and symbolic \lstinline{solve(b)}{}: Due to saving of the whole decomposition object, \dcocppeigen{} achieves a faster run time than Stan.}
\end{wrapfigure}
Referring to the insight gained in Section~\ref{ssec:symImpl}, symbolic evaluations have the potential to drastically improve the performance of AD application. In the case of the matrix-matrix product, actual computations are made using the passive data type and profit from all optimizations in Eigen, while only two additional matrix-matrix products need to be evaluated in the adjoint run. This explains why \dcocppeigen{} as well as the implementation in the \lstinline{multiply}{} function of Stan drastically outperform all other tools.
\par
Since Stan provides optimized linear algebra functions using Eigen, another benchmark was performed for solving a dense system. Stan offers a \lstinline{mdivide_left(A,b)}{} function to solve a system of linear equations which internally will always use the \lstinline{ColPivHouseholderQR}{} decomposition. Therefore, the algorithmic and the \dcocppeigen{} measurements displayed in Fig.~\ref{fig:tools_solver} also utilize this solver class. While Stan uses the same symbolic evaluation from Equations~(\ref{eq:adjointsB})-(\ref{eq:adjointsA}), it performs another decomposition in the adjoint run. \dcocppeigen{} on the other hand keeps the decomposition from the augmented primal run in memory and reuses it later. While Stan keeps the AAD run time overhead \dcocppeigen{} at \( \mathcal O (n^3) \), the implementation in \dcocppeigen{} improves it to \( \mathcal O (n^2) \).

\section{Conclusion \& Outlook}
\label{sec:ConclusionOutlook}
In this work, we have have outlined challenges which occur when calculating derivatives for linear algebra operations using an AD-O tool with the Eigen library. To overcome these issues, the modified library fork Eigen-AD was developed, aimed at authors of AD-O tools to help them improve the performance of their software when applying it to Eigen. Changes to the Eigen source code were kept generic and entry points are provided for a general Eigen-AD base module which can be utilized by an individual AD-O tool module via a dedicated API. Care was taken to realize the improvements via \cpp{} specializations, which keep the look and feel of plain AD-O. General performance improvements were made regarding the usage of expression templates by the AD-O tool and specific operations can now be reimplemented conveniently by an AD-O tool module in order to provide symbolic implementations.
\par
As a showcase, such a module has been implemented for the AD-O tool \dcocpp{}, where important linear algebra operations like the matrix-matrix product or solving of a linear system are differentiated symbolically. Benchmarks have validated the theoretical considerations and underlined the improvements in computational complexity regarding run time and memory usage. It was shown that AD-O tool modules can cancel the AAD overhead for dense solvers with a corresponding implementation and comparisons with other AD-O tools were made to put the produced results into context which further confirmed the improvements.
\par
Eigen-AD is publicly available\footnote{\url{https://gitlab.stce.rwth-aachen.de/stce/eigen-ad}} and other AD software authors are welcome to provide a module for their AD-O tool which can be included in the fork as well as participate in the future development. Investigation into more parts of Eigen are planned in order to extend the Eigen-AD API. Furthermore, there has been communication with the Eigen development team and best efforts were made to keep changes to the Eigen source as general as possible. In combination with the modular setup regarding the Eigen-AD base module and individual tool modules, a partial integration of the changes into future Eigen versions should be discussed. 
\par
All in all, this work has shown the potential of adjusting a linear algebra library to optimize the evaluation of derivatives using an AD-O tool. In the case of Eigen, relatively small changes to its source code allow to provide a general API which can be utilized by other AD-O tools and provide a superior performance compared to ordinary AD-O.

\newpage

\bibliographystyle{splncs04}
\bibliography{references/references}

\end{document}